# Particle-in-cell simulations of anomalous transport in a Penning discharge

## Johan Carlsson[1,4*], Igor Kaganovich[1], Andrew Powis[2], Yevgeny Raitses[1], Ivan Romadanov[3] and Andrei Smolyakov[3]

*[1]Princeton Plasma Physics Laboratory, Princeton, New Jersey 08543, U.S.A.*

*[2]Department of Mechanical and Aerospace Engineering, Princeton University, Princeton, New Jersey 08540, U.S.A.*

*[3]Department of Physics and Engineering Physics, University of Saskatchewan, Saskatoon, Saskatchewan S7N 5E2, Canada*

*[4]Crow Radio and Plasma Science, Princeton, New Jersey 08540, U.S.A.*

**\*Correspondence:** carlsson@pppl.gov

## Abstract

Electrostatic particle-in-cell simulations of a Penning discharge are performed in order to investigate azimuthally asymmetric, spoke-like structures previously observed in experiments. Two-dimensional simulations show that for Penning-discharge conditions, a persistent nonlinear spoke-like structure forms readily and rotates in the direction of ExB and electron diamagnetic drifts. The azimuthal velocity is within about a factor of two of the ion acoustic speed. The spoke frequency follows the experimentally observed scaling with ion mass, which indicates the importance of ion inertia in spoke formation. The spoke provides enhanced (anomalous) radial electron transport, and the effective cross-field conductivity is several times larger than the classical (collisional) value. The level of anomalous current obtained in the simulations is in good agreement with the experimental data. The rotating spoke channels most of the radial current, observable by an edge probe as short pulses.

## I    Introduction

Crossed-field, also known as Hall or ExB, plasma devices are characterized by perpendicular electric ($E$) and magnetic ($B$) fields. Electrons are typically magnetized and ions are unmagnetized, or weakly magnetized. Such devices are relevant to materials processing [1], spacecraft propulsion[2] and as ion beam sources [3]. A commonality among a wide variety of crossed-field devices is the significance of anomalous electron transport across the applied magnetic field and the spontaneous emergence of large-scale structures[4,5,6,7,8,9,10,11]. An understanding of these effects is important for both academic (fundamental plasma physics) and practical (device performance, efficiency and lifetime) purposes.

The Penning discharge is an archetypical crossed-field configuration that exhibits essential features and characteristics of many widely used ExB devices. The device typically features an approximately uniform axial confining magnetic field and plasma is generated at the center of this trap by a source of ionizing electrons.

The preferential diffusion of the weakly magnetized ions results in a radial ambipolar field and therefore ExB drifts in the azimuthal direction. The Penning discharge has similar plasma parameters, electric and magnetic fields as the near-anode region of a Hall thruster, where a spoke-like coherent structure is commonly observed [12,13,14,15,16]. Penning discharges provide more convenient access for diagnostics than Hall thrusters do, and the physics is somewhat simplified by the lack of a magnetic field gradient, eliminating one energy source able to drive instabilities[17].

Early investigations into electron transport within a Penning discharge revealed that the onset of anomalous effects coincided with the formation of an m=1 flute instability[6] with a frequency between 25-50 kHz. This structure was similar in characteristic to the spoke that, at the time, had been observed in Hall thrusters[4]. More recent investigations have been carried out in the Mistral device at the University of Marseille [13]. It was observed that by modifying the source potential, the mode could be transitioned from an m=2, to m=1 to a purely turbulent spectrum [14]. The low mode number structures were correlated strongly with anomalous transport [15], and tomography measurements showed that the structure roughly retained its shape throughout rotation [16]. Further work has been carried out within a Penning discharge at the Princeton Plasma Physics Laboratory (PPPL). Parameter scans with magnetic field and discharge gas showed that the spoke frequency scales as $(B/M)^{0.5}$, where M is the ion mass[12].

Originally it was suggested that the spoke instability was a type of ionization wave[4,5], similar to the concept first proposed by Alfvén [18,19]. It was later realized, however, that the spoke is related to long-wavelength density-gradient-driven flute modes [15,17,20], that saturate at large amplitude. These modes are related to the Simon-Hoh instability[21,22] that occurs when the electric field and density gradient of the plasma are aligned, so that ions are confined by the radial electric field, and density perturbations evolve in such a way that the resulting charge separation enhances the perturbation, leading to instability. The spoke formation is also facilitated by an inverse cascade of energy from higher wavenumber flute modes[17]. Plasma in a Penning discharge is weakly collisional and the Modified Simon-Hoh Instability[21,22] (MSHI) is considered the most likely candidate mode responsible for spoke fluctuations. The MSHI is a collisionless version of the Simon-Hoh instability, where the ion drift velocity (which is much reduced compared to that of the electrons due to the large ion Larmor radius in a weak magnetic field) is taken into account [23,24].

To capture the effects of anomalous transport within numerical models, it is important that simulations be self-consistent, where kinetic effects are important[2]. This makes the Particle-in-Cell technique combined with a Monte-Carlo Collision model (PIC-MCC) well suited for numerical investigations of anomalous transport in crossed-field plasmas. PIC-MCC modeling is computationally expensive, since all time and space scales of the plasma must be resolved, with the electron plasma frequency, or electron cell transit time, and Debye length generally being the most restrictive. This makes it particularly challenging to observe long time and large scale structures. To alleviate this issue, problem models are often scaled for either reduced system size [25] or reduced plasma density or relative permittivity[26].

Improvements in modern computing have led to a number of recent efforts to model Hall plasmas in two dimensions [27,28], and even entire three-dimensional devices [25,29,30,31]. Few simulations have focused on the modeling of a Penning discharge, which has the advantage of being well represented by a two-dimensional simulation model with the axial direction numerically unresolved. These features make the Penning discharge an interesting and potentially more tractable model for exploring anomalous transport and electric self-organization, as well as an attractive target for code validation in preparation for three-dimensional Hall-thruster simulations.

In this paper we base our model on the Penning discharge at PPPL[12]. This experiment has a radius of about 10 cm and length of around 40 cm. It typically operates with axial magnetic field strengths between 10-1000 Gauss with pure argon, xenon or mixtures of the gases giving background gas pressures between 0.1-1 mTorr. The cylindrical device wall is grounded, acting as the anode and a plasma cathode injects a beam of electrons along the axis, resulting in a predominantly radial electric field. A schematic of the device is shown in Fig. 1. Electron density $n_e$ is around $10^{11}$ cm$^{-3}$ and temperature $T_e$ is 1-10 eV. Under a wide range of plasma parameters a rotating, spoke-like coherent structure is observed in experiments, with frequency in the 10s of kHz range.

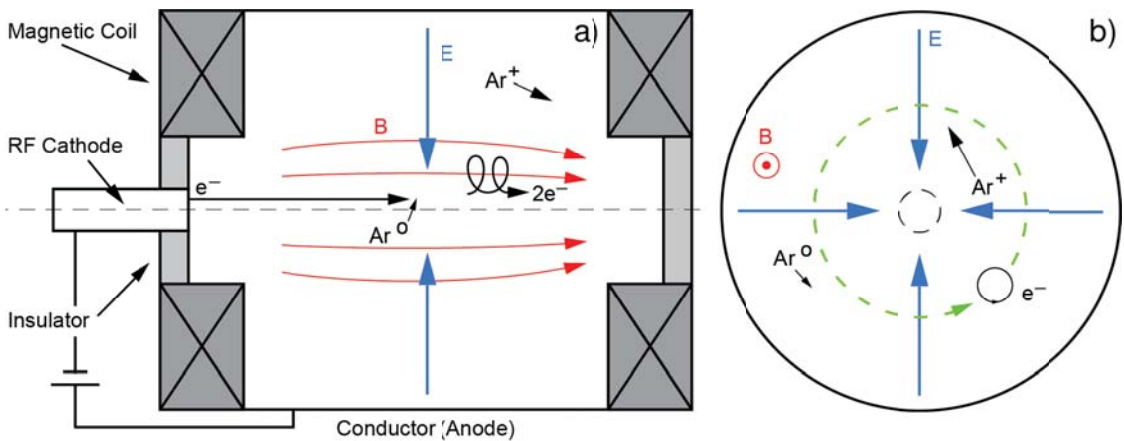

Figure 1: Schematic of the PPPL Penning discharge experiment, a) side view and and b) end view

## II       Simulation model

We use the PIC-MCC code Large-Scale Plasma (LSP)[32], which was recently modified, benchmarked and validated for a low-temperature plasma application[33]. We implemented new Poisson solvers in LSP after being unable to achieve acceptably small residuals with the available solvers. For the simulations reported here, direct solvers were used to solve the discretized Poisson equation. More recently we have also implemented a multigrid Poisson solver to enable scalable three-dimensional simulations of a Hall thruster. Our modifications of LSP have been significant enough to warrant the use of the name PPPL-LSP for our version of the code.

The goal of the work presented here is to perform two-dimensional simulations reproducing the essential physics and main features of spoke formation in the PPPL Penning-discharge experiment and similar devices. Some initial simulations were performed with a one-dimensional model to develop a correctly configured input file with parameters resulting in numerically converged solutions. In the one-dimensional simulation model the plasma is assumed to be azimuthally symmetric and only the (radial) distance from the device axis is numerically resolved. As expected, the one-dimensional model does not reproduce the characteristic features of the experiment, but some nevertheless interesting results are presented in Appendix A. In the two-dimensional simulations, the plane perpendicular to the device axis (i.e. the azimuthal and radial directions) is numerically resolved using Cartesian coordinates. The simulation domain thus corresponds to the end view of the actual device, shown in Fig. 1b. It is expected that azimuthal perturbations are important for the radial anomalous transport. A uniform Cartesian mesh is chosen, despite the cylindrical geometry of the device and the strong on-axis density peaking. Use of a uniform Cartesian mesh helps avoiding potential numerical problems often associated with nonuniform meshes, particularly at the center of the trap. The system axial direction (along the direction of the magnetic field) is not resolved and therefore the spoke structure is here assumed to be uniform in the axial direction.

The device and plasma parameters used for the simulations are chosen as similar as practically possible to those of the actual device. Simulations are performed with either helium, argon or xenon at gas pressures around 200 $\mu$Torr and with an initially quasineutral plasma with 10 eV electron temperature, room-temperature (25.2 meV) ions and densities of $5 \times 10^{11} cm^{-3}$. The resulting Debye length is about 33 $\mu m$. A PIC simulation using the actual device and plasma parameters would then need about $10^4$ cells in each direction to achieve a numerical resolution high enough to avoid spurious grid heating. Such a simulation is feasible in two dimensions, but would fall well into to the territory of capability computing (using the maximum computing power to solve a single large problem in the shortest amount of time), as opposed to the capacity computing (large number of moderately sized simulations) needed for an exploratory physics study. As is often the case for PIC, one must therefore use parameter scaling to reduce the simulation size. The parameter of primary interest is the size of the simulation domain normalized to the Debye length. There are several ways to scale this ratio. The Debye length can e.g. be increased by scaling down the electron density or by scaling up the relative permittivity. For most of our simulations, we are instead using the actual, physical Debye length and simply scale down the device size, typically to 1 cm diameter. To roughly maintain the device size normalized to gyro radius, we compensate by scaling up the magnetic field. We also present some simulation results for a more realistic device size, with the Debye length scaled up using an artificially large relative permittivity.

Similar to the experimental set up, in the simulations a narrow electron beam is injected at the origin (x=0, y=0) along the axis of the device, parallel to the external, homogeneous, axial magnetic field (see Fig. 1). The beam energy is 10 eV with zero thermal velocity. The injection rate is held constant throughout the simulation. A typical injection region is a 10 by 10 cell square, but the simulation results were found to depend only weakly on the exact beam width. This beam represents electrons streaming from the cathode that are accelerated by the axial electric field at the exit of the cathode. The strength of the magnetic field is chosen to magnetize the electrons (with the device radius of 0.5 cm corresponding to about five electron gyro radii), but to leave the ions unmagnetized. A typical value used is 100 Gauss. The simulated electrons undergo anisotropic elastic collisions against the neutral gas atoms and the ions experience charge-exchange collisions with the same[33]. The 10 eV energy of the injected electrons is below the ionization threshold. With no other ion sources, the only ions in the simulation are those initially present. The electrostatic approximation is assumed valid and the Poisson equation is solved to compute the predominantly radial electric field that develops to drive the electron current across the magnetic field. This electric field confines ions, and leads to a peaked plasma density profile.

Because the injected electrons have zero perpendicular thermal energy, there will inevitably be grid heating in a PIC simulation. The cell size was initially set to 20 $\mu m$, but the beam electrons were found to gain a few eV of perpendicular thermal energy during the simulation due to grid heating. This is not believed to significantly affect the results, but the cell size was nevertheless reduced to 10 $\mu m$. The time step is limited by the cell transit time for electrons and was set to 5 ps for early simulations, and then reduced to 2 ps when the cell size was halved.

## III     Simulation results for reduced device size

Simulations are performed with both a square and an approximately circular, stair-stepped device cross-section. The square boundary was initially used because the direct Poisson solver implemented by us recently in LSP had been verified for this configuration. In both cases the boundary is grounded, perfectly conducting metal, which absorbs impinging particles. Secondary electron emission is neglected. The shape of the boundary is found not to have a significant impact on the main simulation results. An 80 mA, 10-eV electron beam is injected along the axis, with homogeneous current density for $-0.01 cm \leq x, y \leq 0.01 cm$ and zero outside. A peaked density profile is formed, which is quasi-neutral, except in the sheath at the wall.

The simulations using a square boundary are resolved by 500x500 cells and a 5 ps time step. The latter simulations with an approximately circular, stair-step boundary use 1000x1000 cells and a 2 ps time step. 25–100 macro particles per species per cell are launched. The global domain is divided into 256 local domains, keeping both field and particle data local to a process. Interprocess communication is done using MPI. Load balancing is done differently in the x and y directions. The global domain is initially divided into 16x16 local domains. In the x direction, cells are redistributed, as needed, among local domains every 50 ns of simulated time by moving the local domain boundaries. In the y direction, the global domain is divided into 16 regions, which are narrower close to the origin, where the density will be the highest. These region boundaries remain fixed throughout the simulation and load balancing is achieved by redistributing processes between regions. I.e., a region close to the axis might be divided in the y direction by more than 16 processes at the end of the simulation, and a region far from the axis by fewer than 16 processes. The redistribution of processes between regions is done every 200 ns, if needed. Running on 256 cores, simulations typically take 5–10 days to execute. Strong scalability is very close to ideal going from 144 to 256 processes, and execution time could thus likely be reduced by running on more cores. However, the increase in queue time when more cores are requested from the scheduler is expected to largely negate that gain.

### III A:    Square device cross section

Fig. 2 shows the density of beam electrons from a simulation with a square device cross section with 200 $\mu$Torr helium gas and a 100 Gauss magnetic field. As can be seen, a spoke-like structure is rapidly formed and persists throughout the remainder of the simulation. With no source of ions, plasma is gradually lost and the simulations are typically run for around 2 $\mu s$. The color plots in Fig. 2 show the density of the electrons injected from the cathode and do not include the electrons initially present in the simulation.

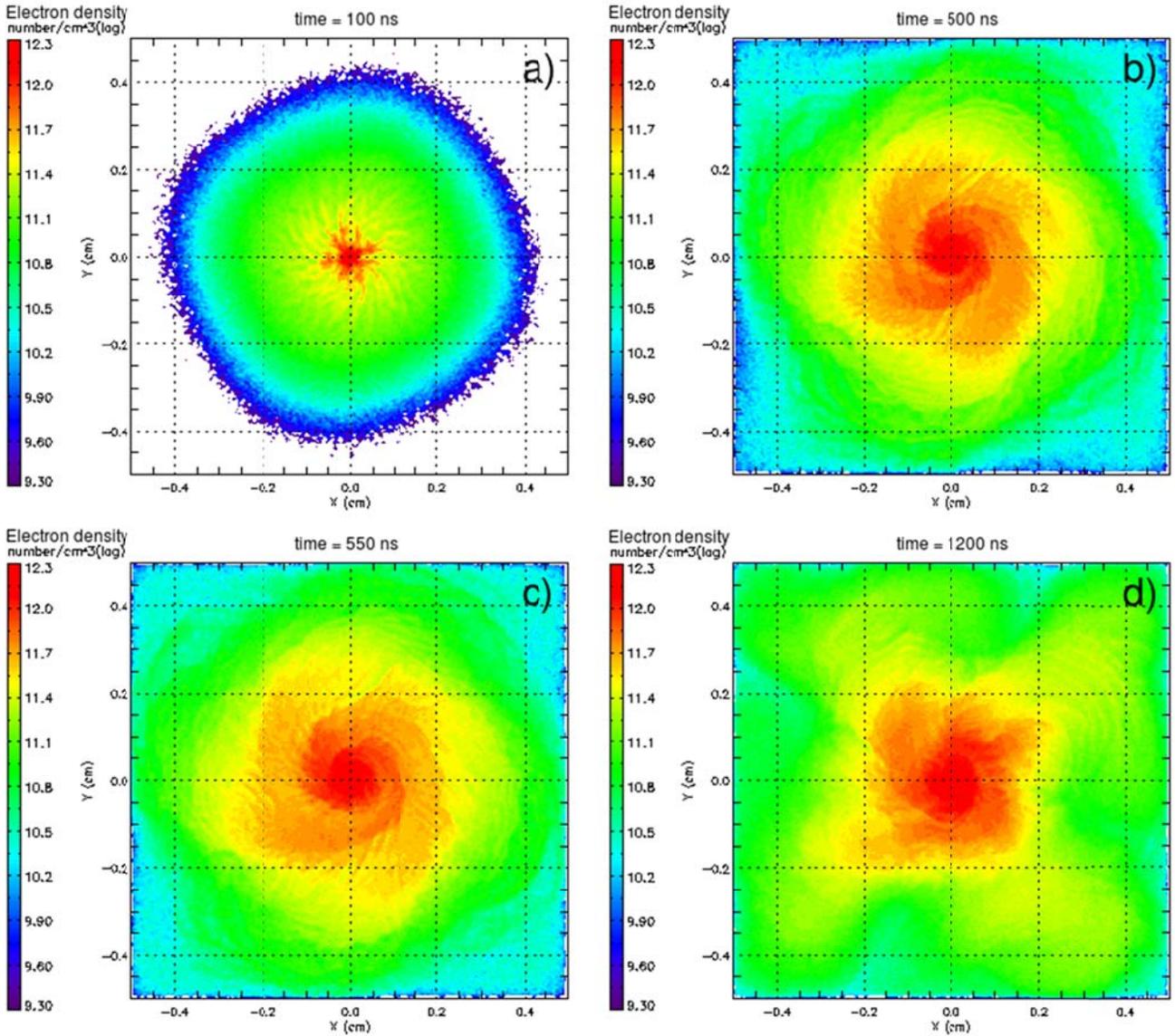

Figure 2: Color plots of density of injected electrons a) 100 ns, b) 500 ns, c) 550 ns and d) 1200 ns after start of injection into an initially uniform plasma with density $5 \times 10^{11} cm^{-3}$ in 200 $\mu$Torr helium gas and a 100 Gauss magnetic field. An animated version of this figure is available[34].

The spoke forms with azimuthal mode numbers in the range 10<$n_\theta$<20, but later settles to smaller mode numbers, $1 \leq n_\theta \leq 4$.

Fig. 3 shows the density profile of beam electrons along the line y=0 at every 10 ns in the time interval 1200–1500 ns. At this stage of the simulation, virtually all of the initially present electrons have been lost at the wall and the density of beam electrons is equal to the total electron density.

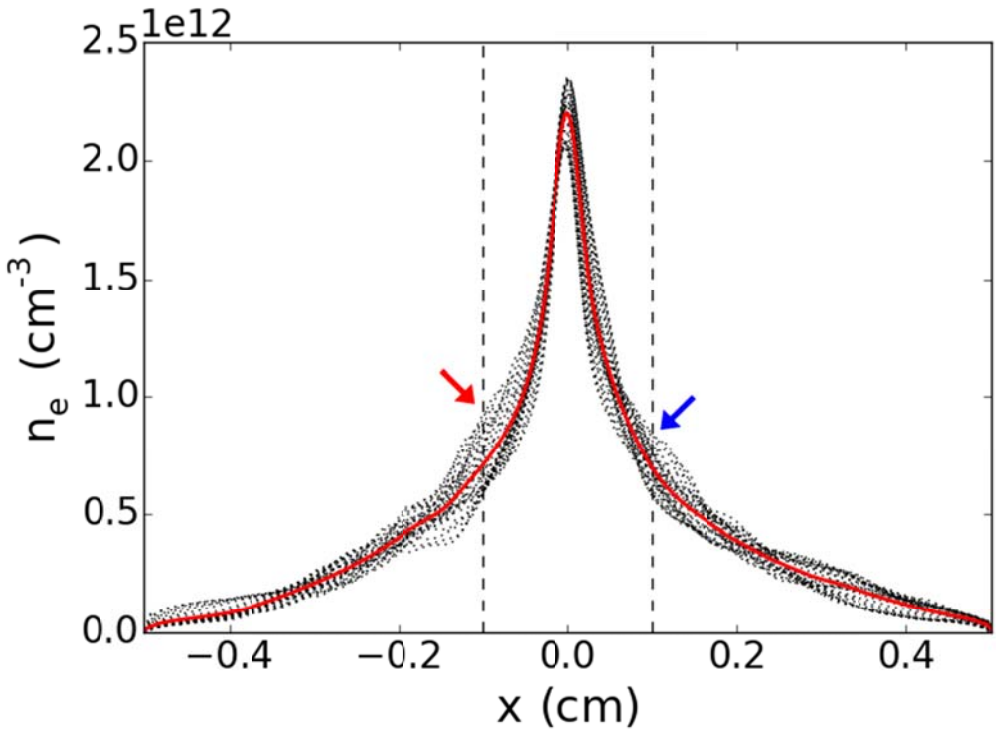

Figure 3: Density profiles of beam electrons along the line y=0 at every 10 ns in the time interval 1200–1500 ns. The dotted graphs are the instantaneous profiles and the solid red graph is their time average. The red and blue arrows indicate the location of two of the synthetic density probes with their signals displayed in Fig. 4. An animated version of this figure is available[35].

As can be seen, the density fluctuations are relatively large and the particle confinement is poor with a potential well that is only a few volts (not shown). To study the density fluctuations in more detail, we plot the local density versus time at $r = 0.1 cm$ at four different azimuthal locations in Fig. 4.

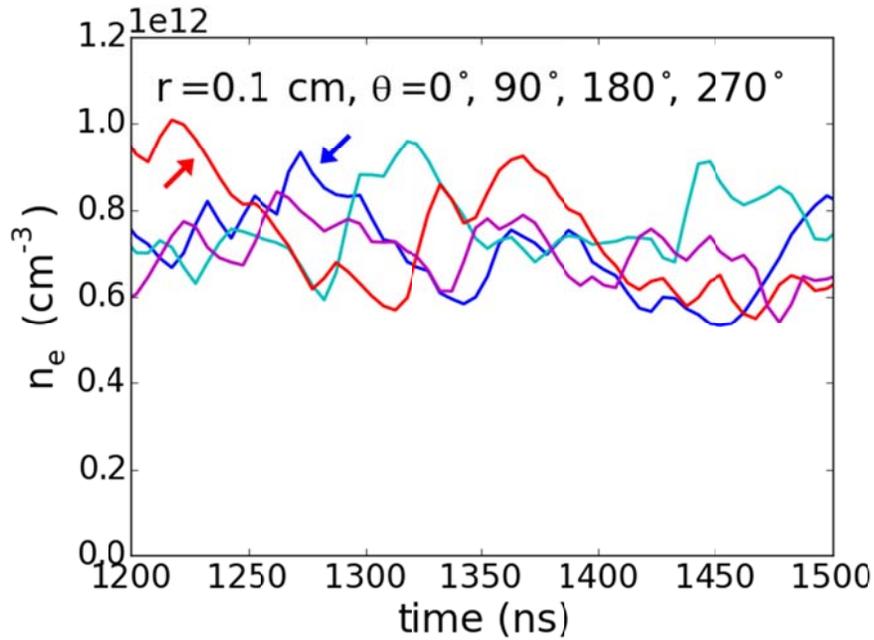

Figure 4: Electron density as function of time at four different azimuthal locations, all at a distance of 0.1 cm from the device axis. Red and blue graphs are for 180° and 0°, cyan and magenta graphs are for 90° and 270°, respectively.

The density time trace presented as the red (blue) graph in Fig. 4 is identical to the density at the location indicated by the red (blue) arrow in Fig. 3. The electron-density amplitude fluctuation at 0.1 cm radius (distance from device axis) is approximately ±20%. The phase difference between the different azimuthal locations corresponds to a rotation in the counter-clockwise direction, which, in our simulation model, corresponds to the ExB and electron diamagnetic drift directions. The rapid emergence of a spoke, rotating in the ExB direction, is a universal feature of our simulations.

Density time traces at a fixed azimuthal location and varying radius are shown in Fig. 5 for a time interval four times longer than that in Fig. 4.

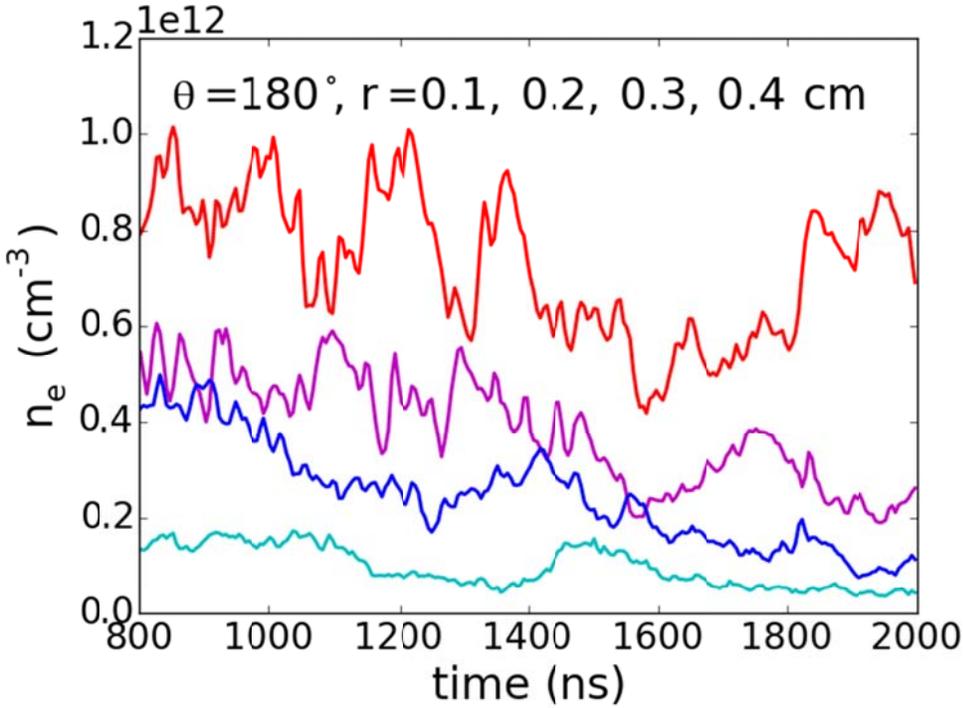

Figure 5: Electron density as function of time at four different radial locations, all at azimuthal angle 180°. Red graph is for radius 0.1 cm (compare with Figs. 3 and 4). Magenta, blue and cyan graphs are for radii 0.2, 0.3 and 0.4 cm, respectively.

The large amplitude of the density perturbation is evident from Fig. 5, where the blue graph (density at 0.3 cm radius) is above the magenta graph (density at 0.2 cm radius) on two occasions. That is, $\partial n_e/\partial r > 0$, as the density gradient locally and intermittently reverses. The relative density perturbation is large (about a factor of two) also at 0.4 cm radius, indicating that the spoke reaches the wall. This is somewhat less evident in Fig. 2, where the logarithmic scale obscures even large perturbations of the low edge density. The only significant difference between ion and the electron densities shown here is that the latter is slightly lagging, creating a potential difference of a few volts (not shown).

On the longer time scale displayed in Fig. 5, it can be deduced that the density oscillation is caused by an approximately rigid-body rotation with a period of around 400 ns, or a frequency of around 2.5 MHz, with higher frequencies prominently present in the vicinity of the device axis. The spoke rotation is in the direction of the ExB and electron diamagnetic drifts (counter clockwise). In the simulations the potential well is shallow, and the two drifts are of similar magnitude for this particular configuration.

To help determine which drift plays a larger role in the anomalous transport in this case, we plot electron-current streamlines over the electron pressure contour (will overlap if diamagnetic drift dominates) and the electrostatic potential (will overlap if ExB drift dominates), respectively. Fig. 6 shows the result after 1.2 $\mu s$ and Fig. 7 after 2 $\mu s$.

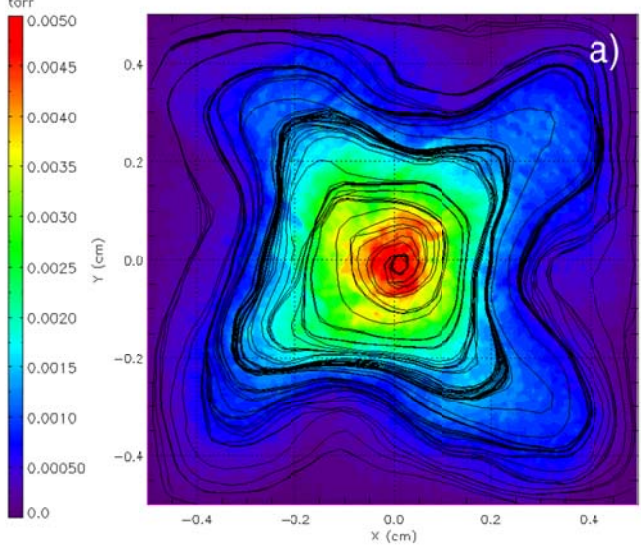 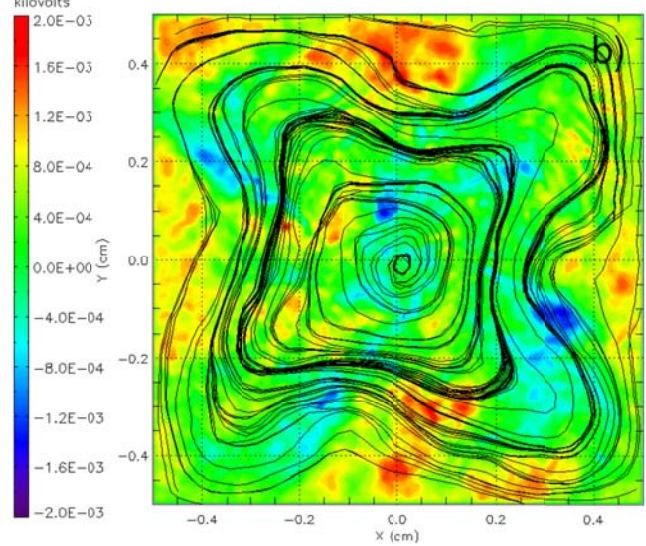

Figure 6: Electron-current streamlines plotted over a) the electron pressure contour and b) the electrostatic potential, respectively, after 1.2 μs.

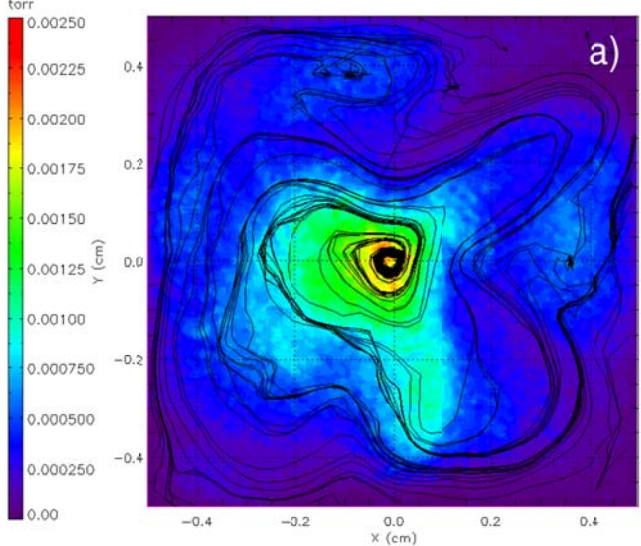 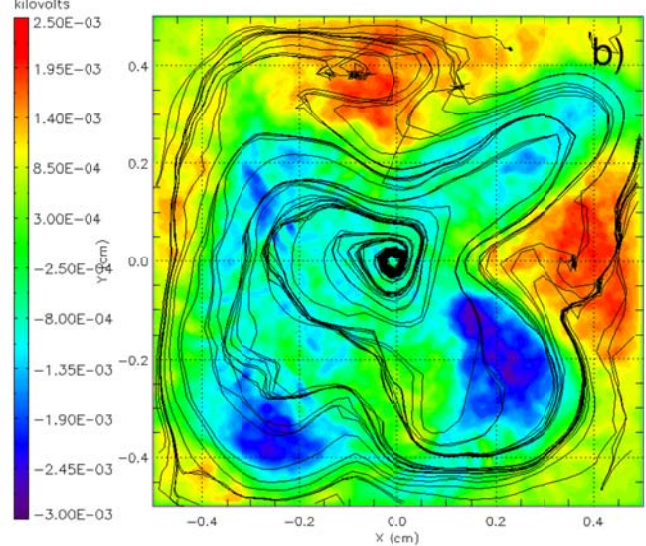

Figure 7: Electron-current streamlines plotted over a) the electron pressure contour and b) the electrostatic potential, respectively, after 2 μs.

As can be seen in Fig. 6, after 1.2 $\mu s$ the electron current follows pressure contours, not potential contours. However, after 2 $\mu s$, the pressure and its gradient have decreased sufficiently for the diamagnetic drift to no longer dominate the ExB drift and the electron current more closely follows potential rather than pressure contours, especially in the edge region, as can be seen in Fig. 7. It therefore seems that neither of the drifts are conclusively dominant. One or the other can be more important, sometimes simultaneously in different parts of the plasma.

## III B: Circular device cross section

To investigate the effect of the shape of the boundary, simulations are performed with the device cross section approximated by a more realistic stair-step circle. To reduce mild grid heating of beam electrons discovered in the simulations with a square boundary, the cell size is also reduced by half, to 10 $\mu m$. To resolve the cell transit time for injected electrons, the time step is correspondingly reduced to 2 ps. Apart from these changes, the simulation model is identical. By comparing Figs. 8 and 2, it can be seen that the main features of the simulation results do not differ for square and circular boundaries: a rotating spoke-like structure rapidly forms, initially with large azimuthal mode numbers, and then settling down with $1 \leq n_\theta \leq 4$. The spoke frequency is also roughly the same for both boundary shapes.

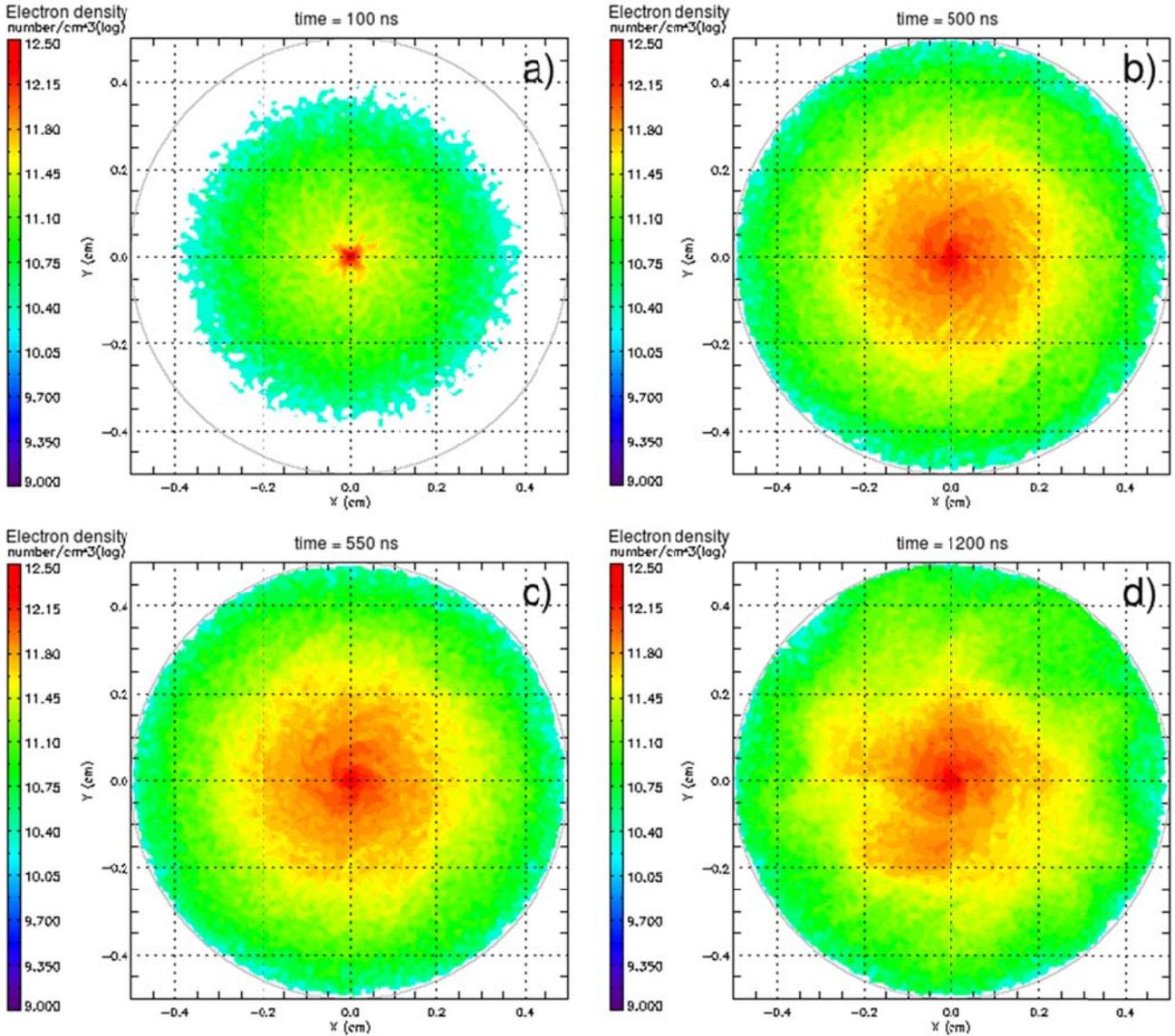

Figure 8: Color plot of density of injected electrons after a) 100 ns, b) 500 ns, c) 550 ns and d) 1200 ns. An animated version of this figure is available [36].

The linear dispersion relation, accounting for density gradient and finite electron Larmor radius, is given by [17, 20]

$$\frac{\omega_* + (\omega - \omega_0 + i\nu_{en})k_\perp^2 \rho_e^2}{\omega - \omega_0 + (\omega - \omega_0 + i\nu_{en})k_\perp^2 \rho_e^2} = \frac{k_\perp^2 c_S^2}{\omega^2},  \qquad (1)$$

where $\omega_*$ is the diamagnetic drift frequency, $\omega_* \equiv k_\theta v_* = -cT_e k_\theta/eBL_n$, $L_n = -n/(dn/dr)$ is the effective density gradient length, $\omega_0 \equiv k_\theta v_0 = cEk_\theta/B$ is the drift frequency, $k_\theta$ is the azimuthal wave vector, $k_r$ is the radial wave number, $k_\perp^2 = k_\theta^2 + k_r^2$, $\rho_e$ is the electron Larmor radius, $c_S$ is the ion sound speed, $\nu_{en}$ is the electron-neutral collision frequency. The following parameters were used for calculations of dispersion relations: $T_e$=5eV, $B = 100G$, $E = 2V/cm$, $n = 5 \cdot 10^{11} cm^{-3}$, $c_S = 5 \cdot 10^5 cm/s$, $v_*$=2· $10^6 cm/s$, $v_0$=2· $10^6 cm/s$, $\omega_{ce}$=2· $10^9 rad/s$, $\omega_{pe}$=5.6· $10^{10} rad/s$, $\rho_e = 0.75\ mm$. The density profile was approximated by $n(r) = n_0 \ln(R/r)$, and correspondingly $L_n(r) = r\ \ln(R/r)$ changes with radius. Wavenumbers are defined as $k_\theta = n_\theta/r$ and $k_r = 2\pi n_r/R$, where R is the system radius. In the limit of very short wavelength $k_\perp^2 \rho_e^2 \gg 1$, the dispersion relation gives $\omega = c_S k_\perp$. For longer-wavelength modes $k_\perp^2 \rho_e^2 \ll 1$ the dispersion relation gives the typical Simon-Hoh instability[17]

$$-\frac{\omega_*}{\omega_0} = \frac{k_\perp^2 c_S^2}{\omega^2}. \qquad (2)$$

And because of our parameters $v_* \approx v_0$, $\omega \approx i c_S k_\perp$, that is the modes growth rate is determined by the ion sound velocity.

The dispersion relation Eq. (1) is consistent with the simulation results, notwithstanding the fact that the spoke is a multimodal, complex nonlinear structure and hence a linear theory that is using axisymmetric density and potential profiles should only be used as rough guidance. As discussed above, azimuthal mode numbers $n_\theta$ with values ranging from 1 to 4 are present in the simulations. Appropriate values for the radial mode numbers $n_r$ are harder to estimate from the simulation results, however the dispersion relation is relatively insensitive to mode-number variations in the reasonable range of $10 \leq n_r \leq 50$. For these parameter values the solution to the dispersion relation is unstable at almost all radii, but the real part of the frequency is an order of magnitude higher than what is observed in the simulations. However, the azimuthal component of the group velocity, which is well approximated by the ion acoustic speed, is within about a factor of two of the spoke velocity, with the latter being higher.

The electron-current streamlines in Figs. 6 and 7 seem to connect with the wall. Fig. 9 shows time traces of edge electron density and radial current, respectively. As can be seen, the radial current is very strongly correlated with the density fluctuations and we can conclude that the spoke carries almost all of the radial leakage current. The edge electron-density azimuthal asymmetry is less visible in Fig. 8 due to its logarithmic density scale, but it is present and in full agreement with Fig. 9.

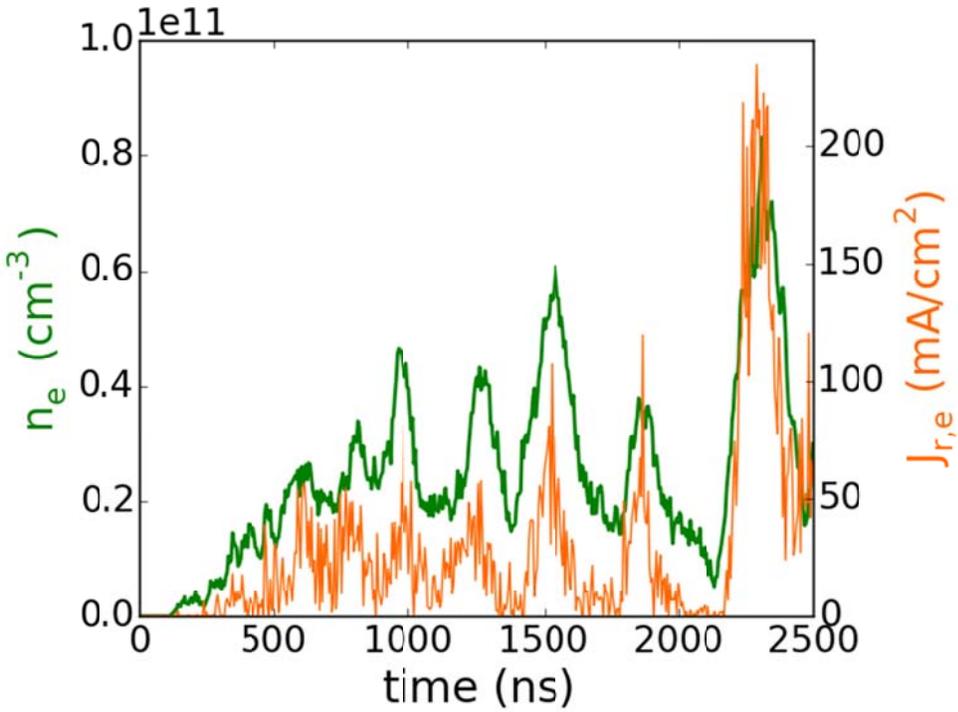

Figure 9: Edge electron density (green) and radial current (orange), both evaluated at (x = 0 cm, y =−0.5 cm), as function of time.

Simulations with argon and xenon show similar results as those with helium. The spoke period, similarly to a trend observed in the experiment, is proportional to $\sqrt{M}$ (see Fig. 10). The heavy-ion simulations are therefore more challenging since they have a slower time scale and require longer simulations.

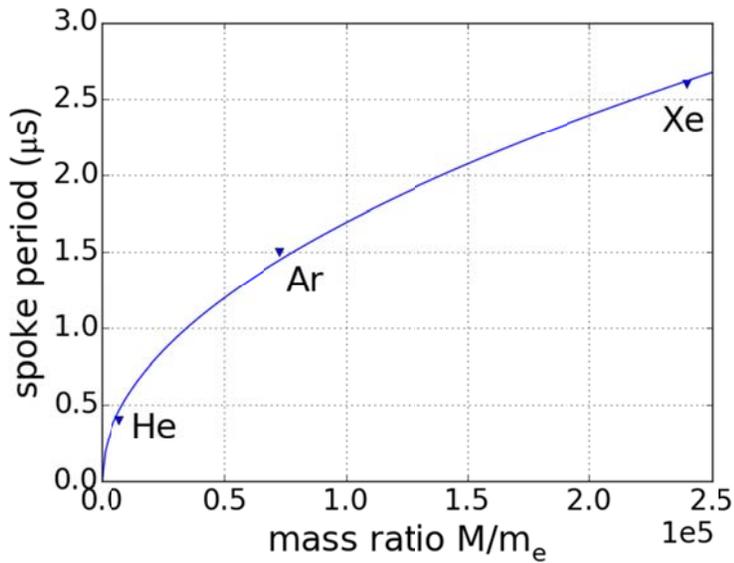

Figure 10: Spoke period vs. ion-to-electron mass ratio. Triangles are approximated values from simulations and solid graph is a fit to a square-root function.

The modes responsible for the anomalous transport are not yet conclusively identified, although the ion-mass scaling of Fig. 10 indicates the importance of ion motion. A phase-space plot of the ions, shown in Fig. 11, exhibits ion dynamics consistent with the charge separation created by the modified Simon-Hoh instability.

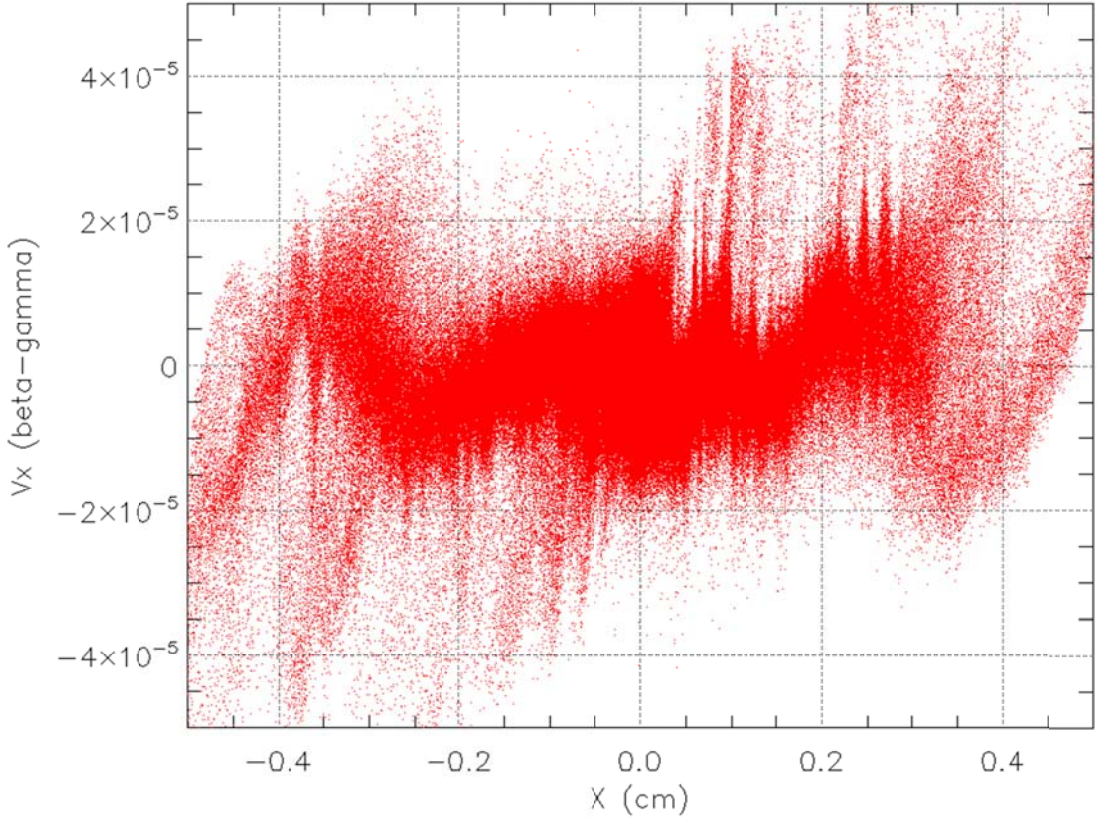

Figure 11: Normalized momentum in x direction vs. x for helium ions in the range −100μm < y < 100μm (compare with density contour of Fig. 8d from same simulation and at same time slice).

A similar plot of electrons in phase space show no discernible structure (not shown).

## IV    Simulation results for realistic device size using increased relative permittivity

One of the candidate unstable modes for generating anomalous transport is the modified Simon-Hoh instability, which is strongly affected by the density gradient [23, 24]. In the simulations with reduced device size it is therefore likely that the Simon-Hoh mode is not accurately reproduced because the density gradients, and therefore mode frequencies, are higher than those observed in experiments[12]. Frequency is approximately inversely proportional to the radius of the device. Moreover, for the small-size system finite electron gyroradius effects become important[17]. The dispersion relation (1) indicates that by scaling the relative permittivity instead of the device size, the Simon-Hoh mode should be unaffected by the scaling. Furthermore, the ion-acoustic mode should be correctly reproduced in scaled-permittivity simulations, and the lower-hybrid mode only slightly modified through its weak dependence on the electron plasma frequency in the relevant parameter regime. We therefore perform simulations with a more realistic device diameter of 5 cm and a gas pressure of 1 mTorr. With a relative permittivity $\varepsilon_r$=100, 500x500 cells are sufficient to resolve the Debye length, which is increased by a factor of 10 by the permittivity scaling. A comparison case with $\varepsilon_r$=400 and doubled cell size and time step, but otherwise identical, is also run.

Fig. 12 shows the density of beam electrons from a simulation with $\varepsilon_r$=100 in a 5 cm by 5 cm simulation domain resolved by 500x500 cells. The results are similar to those with reduced device size, although the structures evolve on a slower time scale.

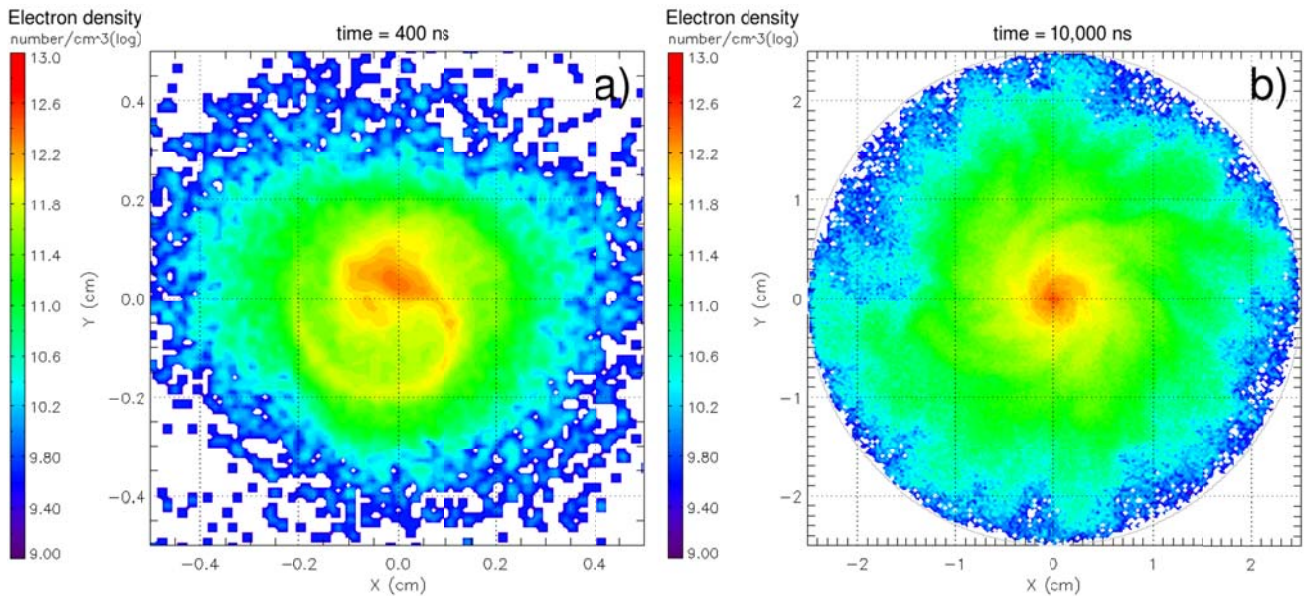

Figure 12: Density of injected electrons with $\varepsilon_r$=100 after a) 400 ns (close up of central region) and b) 10,000 ns.

The beam-electron density contours with $\varepsilon_r$=400 are transiently different than with $\varepsilon_r$=100 [compare Fig. 13(a) with Fig. 12(a)], but then become very similar and remain so throughout the simulation [see Figs. 13(b) and 12(b)].

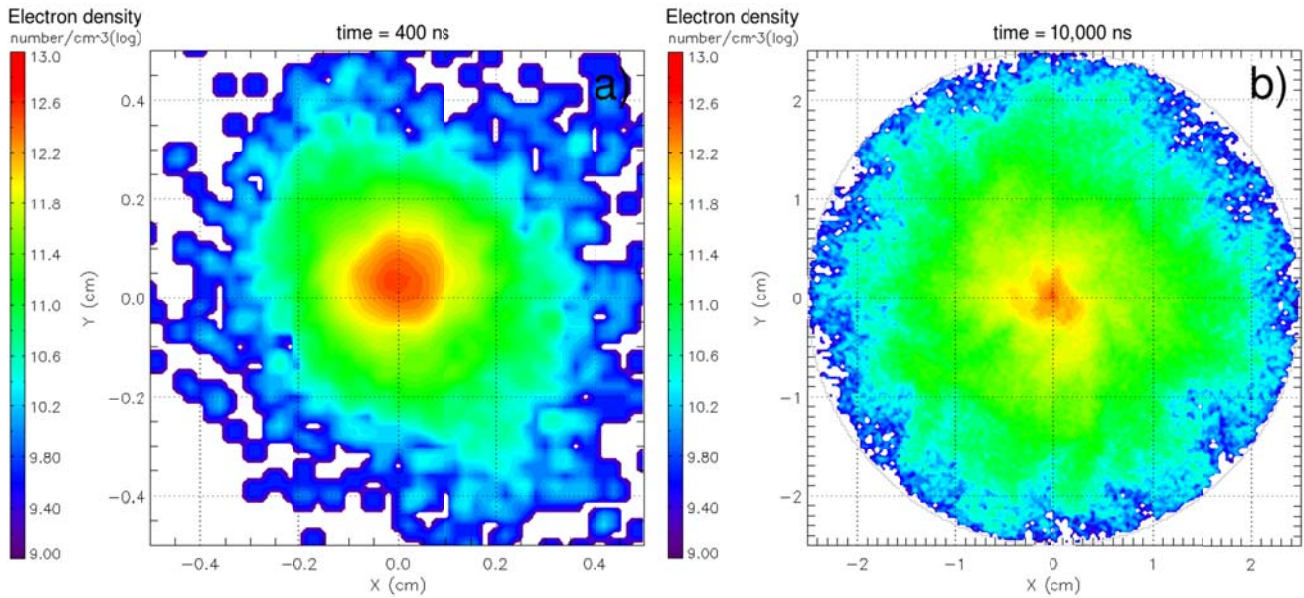

Figure 13: Density of injected electrons with $\varepsilon_r$=400 after a) 400 ns (close up of central region) and b) 10,000 ns, similar to Fig. 12 with $\varepsilon_r$=100.

The Hall parameter, $\omega_{ce}/\nu_{eff}$, is calculated from the simulation results using Ohm's law, neglecting the weak electric field, and is plotted in Fig. 14. Also for the Hall parameter, there is no significant difference between results with $\varepsilon_r$=100 and $\varepsilon_r$=400. The Hall parameter from the simulations is well approximated by the value from the Bohm model for both cases.

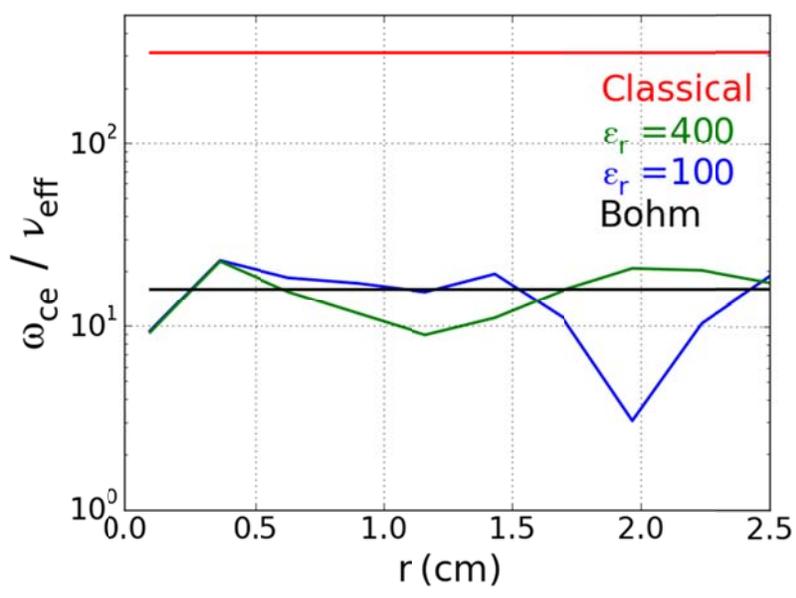

Figure 14: The blue and green graphs show the Hall parameter from simulations as a function of radius for $\varepsilon_r$=100 (blue) and $\varepsilon_r$=400 (green). The red line is the Hall parameter with classical collisions only. The black line is the Hall parameter used in the Bohm model (which has a value of 16).

Fig. 15 shows the edge density and radial-current fluctuation for simulations with $\varepsilon_r$=100. As can be seen, the spoke rotation is far from sinusoidal, although the major frequency component seems to be of the order of 100 kHz, in good agreement with experimental values[12], when scaled for ion mass. In future work, longer simulations will be performed to obtain more conclusive results. For the larger value of relative permittivity, the spoke at the edge seems to be characterized by higher frequencies (not shown). The results for the spoke frequency with scaled permittivity are therefore still only tentative.

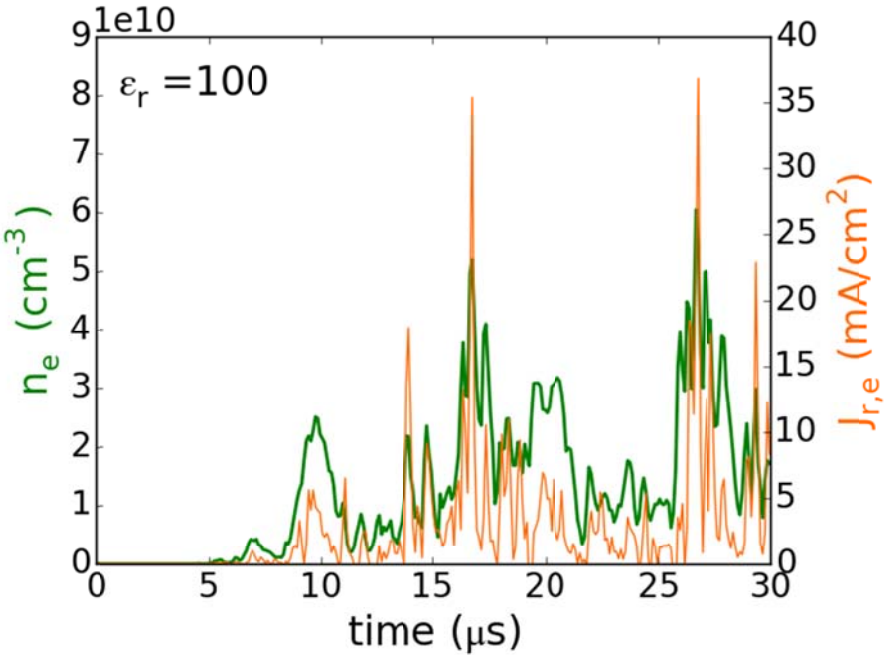

Figure 15: Edge electron density (green) and radial current (orange), evaluated at (x = 0 cm, y =−2.5 cm), as function of time for $\varepsilon_r$=100.

Agreement between the dispersion relation Eq. (1) and the simulations with more realistic device size is relatively good, using parameters estimated from the simulation results. The real frequency of the dispersion-relation solution is again about an order of magnitude higher than the range of frequencies that are prominent in the simulations. The azimuthal component of the group velocity compares better with the simulated spoke rotation speed, with the latter being slightly lower in this case.

## V. Conclusions

We have performed two-dimensional electrostatic particle-in-cell simulations of anomalous transport in an ExB configuration corresponding to a Penning discharge. The electrons injected by an RF cathode in the actual device are modeled as a 10 eV (below the ionization threshold) beam of cold electrons, injected along the device axis. Any effects of ionization are neglected in the simulations presented here, but will be investigated in future work. Preliminary results indicate that the spoke in a Penning discharge is only weakly affected by ionization. The simulations presented here exhibit formation of an azimuthally asymmetric, spoke-like structure rotating in the direction of the ExB and electron diamagnetic drifts with average azimuthal velocity within a factor of two of the dispersion-relation group velocity, which is well approximated by the ion sound speed. In addition to larger structures, small-scale modes also contribute to the anomalous transport. However, the rotating spoke channels most of the radial current in short bursts when the "arc" of the spoke connects a point of the discharge periphery to the center, allowing electrons to move along equipotential lines. The observed spoke frequency is in reasonable agreement with the dispersion relation and scales with the square root of the ion mass similarly to the scaling observed in experimental data. The Hall parameter and corresponding anomalous current obtained in the simulations are also similar to experimental values.

PIC simulations must resolve the Debye radius to avoid numerical heating. If the simulation domain is large compared to the Debye length, PIC simulations can become intractably large. Therefore, simulations were scaled in two ways: either by reducing the size of the simulation domain or by artificially increasing the relative permittivity. The spoke was observed in both cases. With the former scaling (reduced plasma size), the frequency of the mode is higher, more similar to a lower-hybrid mode in accord with the generalized dispersion relation. With the latter scaling (increased permittivity), the frequency of the mode corresponds better to the Simon-Hoh instability.

As expected from the dispersion relation for the modes relevant for anomalous transport, scaling up the relative permittivity does not significantly affect the simulation results. Further investigation of the effects of the artificially increased relative permittivity is underway.

## Supplementary material

Animated versions of Figs. 2, 3 and 8 are provided in the form of five video files. The file PenningSquare-nb-log_vs_xy.mp4 is an animated version of Fig. 2. The file PenningSquare-nb-profile.mp4 is an animated version of Fig. 3. The files PenningCircle-nb-log_vs_xy.mp4, PenningCircleCloseUp-nb-log_vs_xy.mp4 and PenningCircleCloseUpLarge-nb-log_vs_xy.mp4 are animated versions of Fig. 8, the latter two being close ups of the central injection region.

## Appendix A: One-dimensional simulations

It is evident that a one-dimensional simulation model of the Penning discharge, numerically resolving only the radial direction, would be simplistic and unable to reproduce some of the fundamental physics of the anomalous transport in the device. We nevertheless initially performed some one-dimensional simulations, primarily to identify and correct errors in the input file and to find spatial and temporal resolution resulting in numerically converged solutions. In this appendix we will present some physics results from the one-dimensional simulations that are not directly pertinent to a Penning discharge, but illustrate some basic aspects of anomalous transport in magnetized plasmas.

The one-dimensional simulations are performed in a domain $-0.5 cm \leq x \leq 0.5 cm$, bounded by grounded, perfectly conducting metal. A cell size of 10 $\mu m$ is used, a third of the Debye radius, which limits grid heating. The time step is 5 ps. These simulations are unable to reproduce azimuthally asymmetric phenomena, such as the waves and rotating spoke seen in the experiment. With the azimuthal and axial wave numbers forced to zero, the only remaining fluctuations are one-dimensional lower-hybrid modes, while the azimuthal ExB and diamagnetic flows (due to the electric field and density gradient, respectively) are removed as sources of instability. In the one-dimensional case, the anomalous transport is much reduced as only the second order drift velocity contributes to the transport. As a result of the lower value of the anomalous transport and higher electric field that confines ions, the plasma density profile is much narrower for one-dimensional simulations compared to 2D simulations and experimental results.

With a 13 mA electron beam injected around the axis in the region $-0.01 cm \leq x \leq 0.01 cm$, the one-dimensional simulations reach steady state after about 5 $\mu s$. Steady state is reached by balancing the continuous axial current injection with a radial leakage current. A deep potential well develops, not seen in the experiment, which indefinitely confines the ions. The ion energy distribution function, with Hamiltonian (or total) energy as the independent variable, is very nearly Maxwellian and the density is well approximated by the Boltzmann distribution. A quasineutral plasma exists around the axis of the device, surrounded by a very large sheath region with only electrons present. The electrons rotate at the ExB velocity.

To investigate if there is any anomalous transport with the one-dimensional simulation model, we first perform a simulation at zero helium gas pressure, i.e. without classical, collisional transport. The electron density and electrostatic potential are shown in Fig. A1.

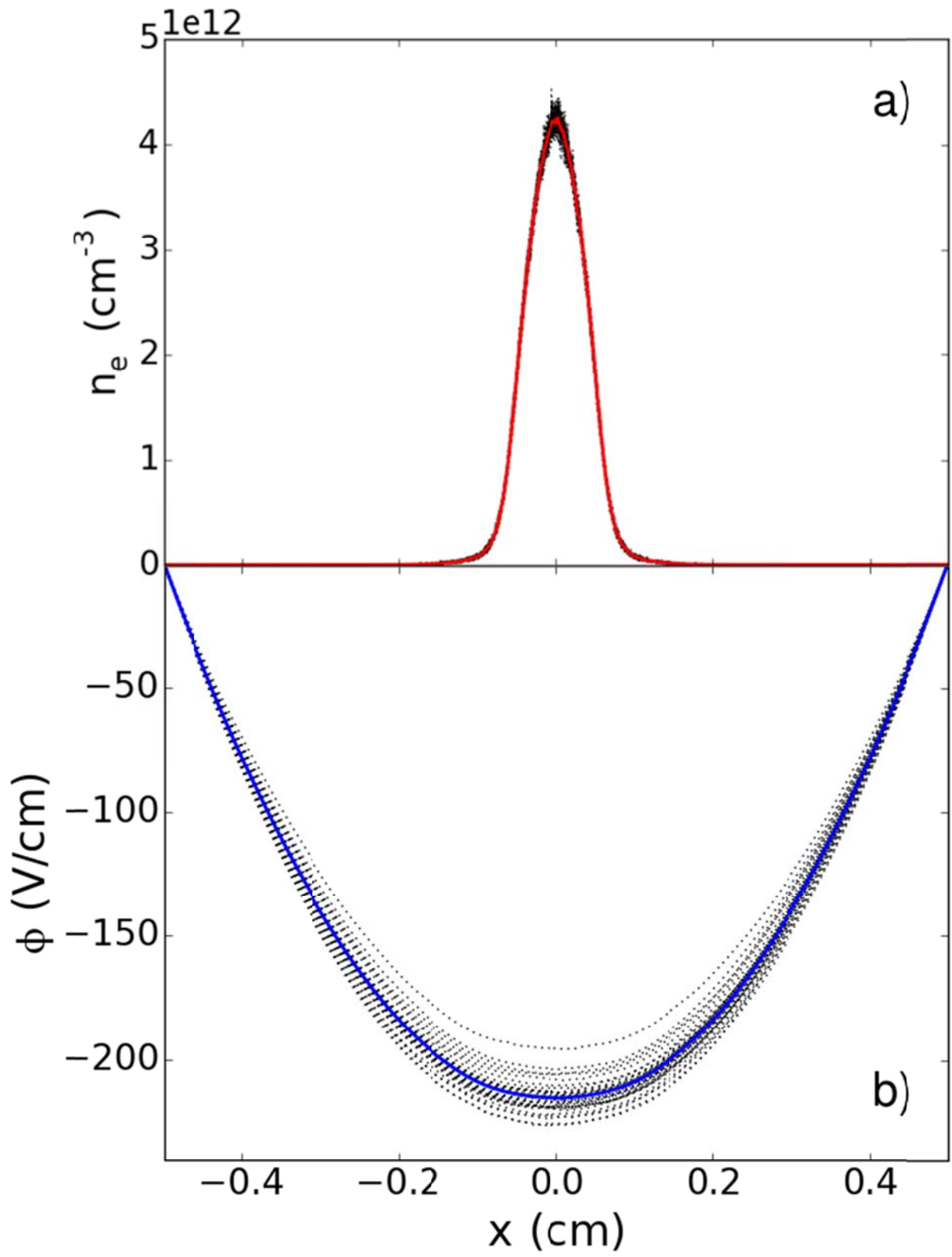

Figure A1: One-dimensional simulation of Penning discharge at zero helium gas pressure with a) electron density and b) electrostatic potential. The dotted graphs are snapshots of the profiles and the solid, colored graphs are time averages.

A number (31) of different snapshots are plotted as dotted graphs to show the fluctuations around the mean. As can be seen in Fig. A1, both electron density $n_e$ and electrostatic potential $\phi$ fluctuate.

A 3.3 MHz oscillation is seen in the simulations, a frequency corresponding to the lower-hybrid mode $\omega_{LH} = (\omega_{ce}\omega_{ci})^{1/2}$. By varying the simulation parameters we find that the mode frequency is proportional to the magnetic field and inversely proportional to the square root of the ion mass, but weakly dependent of the gas pressure. Increasing pressure does however reduce the on-axis density and broadens its profile, and makes the potential well shallower. At 100 $\mu$Torr, a value in the experimental parameter range, the on-axis potential is about -70 V, compared to a few negative volts seen in the experiment. The one-dimensional simulations thus do exhibit anomalous transport, driven by the one-dimensional lower-hybrid modes. In the one-dimensional case the lower-hybrid modes are excited by the radial electron current, which is weak, as only the second order electron drifts contribute to the radial flow.

An effective collision frequency $\nu_{eff}$ can be calculated from the simulation data using Ohm's law with both radial-electric and pressure-gradient terms. Radial profiles of the Hall parameter, the ratio of electron gyro frequency $\omega_{ce}$ and $\nu_{eff}$, from simulations at two different pressures are shown in Fig. A2.

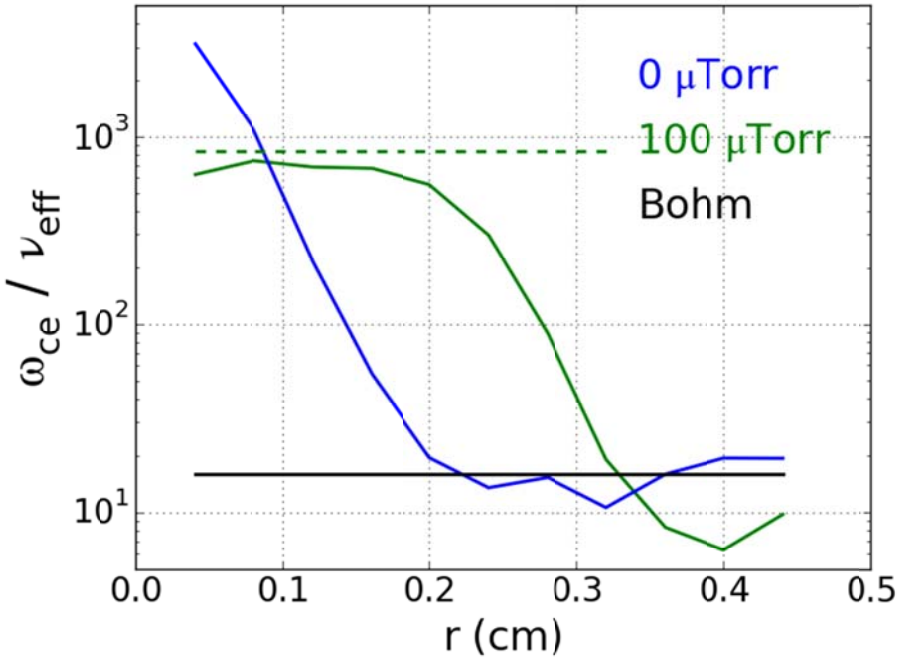

Figure A2: The solid colored graphs show the Hall parameter from 1D simulations as a function of radius. The dashed line is the Hall parameter with classical collisions only. The solid black line is the Hall parameter used in the Bohm model.

With finite gas pressure, the Hall parameter near the axis is close to the classical value, but somewhat lowered by anomalous collisions. Close to the edge, all the simulations are in good agreement with the Bohm model. The values of the Hall parameter near the axis are at least an order of magnitude larger than in the experiment. The discrepancy is due to the different nature of the anomalous transport in the one-dimensional case, which is at least an order of magnitude smaller than in the more realistic two-dimensional case.

## Acknowledgments

This work was supported by the Air Force Office of Scientific Research under grant number FA9550-17-1-0010. This research was enabled in part by support provided by WestGrid (www.westgrid.ca) and Compute Canada / Calcul Canada (www.computecanada.ca). Some of the simulations presented in this article were performed on computational resources supported by the Princeton Institute for Computational Science and Engineering (PICSciE) and the Office of Information Technology's High Performance Computing Center and Visualization Laboratory at Princeton University.

[1] J. T. Gudmundsson, N. Brenning, D. Lundin and U. Helmersson, "High power impulse magnetron sputtering discharge", *Journal of Vacuum Science & Technology A: Vacuum, Surfaces, and Films* **30**, 030801 (2012).

[2] Jean-Pierre Boeuf, "Tutorial: Physics and modeling of Hall thrusters", *Journal of Applied Physics* **121**, 011101 (2017).

[3] Harold R. Kaufman, Raymond S. Robinson and Richard Ian Seddon, "End-Hall ion source", *Journal of Vacuum Science & Technology A: Vacuum, Surfaces, and Films* **5**, 2081 (1987).

[4] G. S. Janes and R. S. Lowder. "Anomalous Electron Diffusion and Ion Acceleration in a Low-Density Plasma", *The Physics of Fluids* **9**, 1115 (1966).

[5] Yu. B. Esipchuk, A. I. Morozov, G. N. Tilinin and A. V. Trofimov, "Plasma oscillations in closed-drift accelerators with an extended acceleration zone", *Soviet Physics Technical Physics* **18**, 928 (1974).

[6] K. I. Thomassen, "Turbulent Diffusion in a Penning-Type Discharge", *The Physics of Fluids* **9**, 1836 (1966).

[7] J. B. Parker, Y. Raitses, and N. J. Fisch, "Transition in electron transport in a cylindrical Hall thruster", *Applied Physics Letters* **97**, 091501 (2010).

[8] C. L. Ellison, Y. Raitses, and N. J. Fisch, "Cross-field electron transport induced by a rotating spoke in a cylindrical Hall thruster", *Physics of Plasmas* **19**, 013503 (2012).

[9] Michael S McDonald, and Alec D. Gallimore, "Rotating spoke instabilities in Hall thrusters", *IEEE Transactions on Plasma Science* **39**, 2952 (2011).

[10] Michael J. Sekerak, Benjamin W. Longmier, Alec D. Gallimore, Daniel L. Brown, Richard R. Hofer and James E. Polk, "Azimuthal spoke propagation in Hall effect thrusters", *IEEE Transactions on Plasma Science* **43**, 72 (2015).

[11] E. Y. Choueiri, "Anomalous resistivity and heating in current-driven plasma thrusters", *Physics of Plasmas* **6**, 2290 (1999).

[12] Y. Raitses, I. Kaganovich and A. Smolyakov, "Effects of Gas Pressure on Low Frequency Oscillations in ExB Discharges", *34th International Electric Propulsion Conference*, IEPC-2015-307.

[13] M. Matsukuma, Th. Pierre, A. Escarguel, D. Guyomarc'h, G. Leclert, F. Brochard, E. Gravier, Y. Kawaia, "Spatiotemporal structure of low frequency waves in a magnetized plasma device", *Physics Letters A* **314**, 163 (2003).

[14] R. Barni, C. Riccardi, Th. Pierre, G. Leclert, A. Escarguel, D. Guyomarc'h and K. Quotb, "Formation of spiral structures and radial convection in the edge region of a magnetized rotating plasma", *New Journal of Physics* **7**, 225 (2005).

[15] S. Jaeger, Th. Pierre, and C. Rebont, "Direct observation of a cross-field current-carrying plasma rotating around an unstable magnetized plasma column", *Physics of Plasmas* **16**, 022304 (2009).

[16] P. David, A. Escarguel, Y. Camenen and R. Baude, "Characterisation of coherent rotating modes in a magnetised plasma column using a mono-sensor tomography diagnostic", *Physics of Plasmas* **23**, 103511 (2016).

[17] A. I. Smolyakov, O. Chapurin, W. Frias, O. Koshkarov, I. Romadanov, T. Tang, M. Umansky, Y. Raitses, I. D. Kaganovich and V. P. Lakhin, "Fluid theory and simulations of instabilities, turbulent transport and coherent structures in partially-magnetized plasmas of discharges", *Plasma Physics and Controlled Fusion* **59**, 014041 (2016).

[18] Hannes Alfvén, "On the cosmogony of the solar system", *Stockholms Observatoriums Annaler* **14**, 2-1 (1942).

[19] Hannes Alfvén, "On the origin of the solar system", *Quarterly Journal of the Royal Astronomical Society* **8**, 215 (1967).

[20] I. Romadanov, A.I. Smolyakov, Y. Raitses, I. Kaganovich, S. Ryzhkov, "Structure of nonlocal gradient-drift instabilities in Hall E × B discharges", *Phys. Plasmas* **23**, 122111 (2016).

[21] Albert Simon, "Instability of a partially ionized plasma in crossed electric and magnetic fields", *Physics of Fluids* **6**, 382 (1963).

[22] F. C. Hoh, "Instability of Penning-type discharges", *Physics of Fluids* **6**, 1184 (1963).

[23] Y. Sakawa, C. Joshi, P. K. Kaw, F. F. Chen and V. K. Jain, "Excitation of the modified Simon-Hoh instability in an electron beam produced plasma", *Physics of Fluids B: Plasma Physics* **5**, 1681 (1993).

[24] Y. Sakawa and C. Joshi, "Growth and nonlinear evolution of the modified Simon-Hoh instability in an electron beam-produced plasma", *Physics of Plasmas* **7**, 1774 (2000).

[25] K. Matyash, R. Schneider, Y. Raitses and N. J. Fisch, "Numerical investigations of a cylindrical Hall thruster", *Proceedings of the 32nd International Electric Propulsion Conference*, 2011.

[26] James Joseph Szabo, "Fully kinetic numerical modeling of a plasma thruster", Diss. Massachusetts Institute of Technology, 2001.

[27] Jean-Pierre Boeuf and Bhaskar Chaudhury, "Rotating instability in low-temperature magnetized plasmas", *Physical Review Letters* **111**, 155005 (2013).

[28] Jean-Pierre Boeuf, "Rotating structures in low temperature magnetized plasmas—insight from particle simulations", *Frontiers in Physics* **2**, 74 (2014).

[29] F. Taccogna and P. Minelli, "Three-Dimensional Fully Kinetic Particle-in-Cell Model of Hall-Effect Thruster", *International Electric Propulsion Conference*, 2011.

[30] Francesco Taccogna, Pierpaolo Minelli and Mario Capitelli, "Physics of Hall-effect thruster by particle model", *AIP conference proceedings*. Vol. 1501. No. 1. AIP, 2012.

[31] K. Matyash, R. Schneider, S. Mazouffre, S. Tsikata, Y. Raitses and A. Diallo, "3D simulation of the rotating spoke in a Hall thruster", *Proceedings of the 33rd International Electric Propulsion Conference*. 2013.

[32] https://www.orbitalatk.com/lsp/downloads_files/lsp_manual.pdf.

[33] Johan Carlsson, Alexander Khrabrov, Igor Kaganovich, Timothy Sommerer and David Keating, "Validation and benchmarking of two particle-in-cell codes for a glow discharge", *Plasma Sources Sci. Technol.* **26,** 014003 (2017).

[34] Supplemental Information to this paper https://www.youtube.com/watch?v=ovzXZeWVOXA

[35] Supplemental Information to this paper https://www.youtube.com/watch?v=Q_SXVSI4f4Q

[36] Supplemental Information to this paper https://www.youtube.com/watch?v=vVUHdZkL0hI